\documentclass[11pt]{cernrep}
\usepackage{graphicx,xspace}
\usepackage[latin1]{inputenc}
\usepackage{cite,./mcite}
\usepackage{epsfig}
\usepackage{xspace}
\usepackage{graphics}

\newcommand{\pythia}{P\scalebox{0.8}{YTHIA}\xspace}

\newcommand{\herwig}{\scalebox{0.8}{HERWIG}\xspace}
\newcommand{\jimmy}{J\scalebox{0.8}{IMMY}\xspace}

\def\dd{\partial}

\def\pom{{I\!\!P}}

\def\to{\rightarrow}

\def\l{\lambda}

\def\({\left(}
\def\){\right)}

\def\citenum#1{{\def\@cite##1##2{##1}\cite{#1}}}
\def\citea#1{\@cite{#1}{}}

\def\l1vt{\vec{l_{1\perp}}}

\def\jol1{$J_0(\,l_{1\perp}\,r_{\perp}\,)$}

\def\beq{\begin{equation}}
\def\eeq{\end{equation}}
\def\bea{\begin{eqnarray}}
\def\eea{\end{eqnarray}}

\def\df2dlnq2{\dd{F_2}/\dd\log{Q^2}}

\vspace{2cm}

\begin{document}
\title{Survival Probability of Large Rapidity Gaps}

\author{E.~Gotsman, E.~Levin, U.~Maor, E.~Naftali, A.~Prygarin}
\institute{HEP Department, School of Physics and Astronomy, Raymond
  and Beverly Sackler Faculty of Exact Science, Tel Aviv University,
  Tel Aviv, 69978, ISRAEL.} \maketitle
 
\begin{abstract}
Our presentation centers on the consequences of s-channel unitarity, 
manifested by soft re-scatterings of the spectator
partons in a high energy diffractive process, focusing on 
the calculations of gap survival probabilities.
Our emphasis is on recent estimates relevant to exclusive 
diffractive Higgs production at the LHC.
To this end, we critically re-examine the comparison of 
the theoretical estimates of 
large rapidity gap hard di-jets 
with the measured data, and  
remark on the difficulties in the interpretation of HERA 
hard di-jet photoproduction.
\end{abstract}

\section{Introduction}

A large rapidity gap (LRG) in an hadronic, photo or DIS induced  
final state is experimentally defined as a large gap in the 
$\eta-\phi$ lego plot devoid of produced hadrons. 
LRG events were suggested\cite{Dokshitzer:1987nc,Dokshitzer:1991he,
Bjorken:1991xr,Bjorken:1992er} as a 
signature for Higgs production 
due to a virtual $W-W$ fusion subprocess. 
An analogous pQCD process, 
in which a colorless exchange ("hard Pomeron") replaces the 
virtual W, has a considerably larger discovery potential as it 
leads also to an exclusive $p+H+p$ final state. 
Assuming the Higgs mass to be in the range of $100-150\,GeV$, the 
calculated rates for this channel, utilizing proton tagging are 
promissing. 
Indeed, LRG hard di-jets, produced via the same production mechanism, 
have been observed in the 
Tevatron\cite{Abachi:1994hb,Abachi:1995gz,Abbott:1998jb,Abe:1994de,Abe:1997ie,
Abe:1998ip,Affolder:2000vb,Affolder:2000hd,Goulianos:2003ik,
Goulianos:1999wi,Goulianos:2000gp,Goulianos:2004as,Dino} 
and HERA\cite{Derrick:1993xh,Derrick:1995wv,Derrick:1995tw,
Derrick:1995pb,Breitweg:1998gc,Ahmed:1994nw,Ahmed:1995ns,
Adloff:1997sc,schilling,Savin:2002eg,Abramowicz:2004dt,gutsche}. 
The experimental LRG di-jets production 
rates are much smaller than the pQCD (or Regge) estimates. 
Following Bjorken\cite{Bjorken:1991xr,Bjorken:1992er}, 
the correcting damping factor is called "LRG survival probability". 

The present summary aims to review and check 
calculations of the survival probability as applied to the 
HERA-Tevatron data and explore the consequences for 
diffractive LRG channels at LHC with a focus on
diffractive Higgs production.

We distinguish between three 
configurations of di-jets (for details see Ref.\cite{Goulianos:2003ik,Goulianos:1999wi,Goulianos:2000gp,Goulianos:2004as,Dino}):
\newline
1) A LRG separates the di-jets system from the other 
non diffractive final state 
particles. On the partonic level this is a single diffraction (SD) 
Pomeron exchange process denoted GJJ.
\newline
2) A LRG separates between the two hard jets. This  
is a double diffraction (DD) denoted JGJ.
\newline
3) Centrally produced di-jets are separated by a LRG on each 
side of the system. This is a central diffraction 
(CD) two Pomeron exchange process denoted GJJG. 
This mechanism also leads to diffractive 
exclusive Higgs production.

We denote the theoretically 
calculated rate of a LRG channel by $F_{gap}$. 
It was noted by Bjorken\cite{Bjorken:1991xr,Bjorken:1992er}
that we have to distinguish between 
the theoretically calculated rate and the actual measured 
rate $f_{gap}$  
\begin{equation}\label{1.1}
f_{gap}\,=\,\langle{\mid S \mid}^2\rangle \cdot F_{gap}.
\end{equation}
The proportionality damping factor\cite{Gotsman:1992ui,Gotsman:1993ux,Gotsman:1995gg,Gotsman:1994gc} is the 
survival probability of a LRG.
It is the probability of a given LRG not to be 
filled by debris (partons and/or hadrons). 
These debris originate 
from the soft re-scattering of the
spectator partons resulting in a survival probability denoted
${\mid S_{spec}(s) \mid}^2$,
and/or
from the gluon radiation emitted by partons
taking part in the hard interaction with a corresponding survival
probability denoted
${\mid S_{brem}(\Delta y) \mid}^2$, 
\begin{equation}\label{1.2}
\langle{\mid S(s,\Delta y) \mid}^2\rangle\,=\,
\langle{\mid S_{spec}(s) \mid}^2\rangle \cdot
\langle{\mid S_{brem}(\Delta y) \mid}^2\rangle.
\end{equation}
$s$ is the c.m. energy square of the colliding particles and 
$\Delta{y}$ is the large rapidity gap.
Gluon radiation from the interacting partons is strongly suppressed by the 
Sudakov factor\cite{Forshaw:proc}. 
However, since this suppression is included in the 
perturbative calculation (see {\bf 4.3}) we can neglect 
$\langle{\mid S_{brem}(\Delta y) \mid}^2\rangle$ in our calculations. 
In the following we denote 
$\langle\,{\mid S_{spec} \mid}^2\rangle\,=\,S^2$.
It is best defined in impact parameter space (see 
{\bf{2.1}})). 
Following Bjorken\cite{Bjorken:1991xr,Bjorken:1992er}, the survival probability
is determined as the normalized
integrated product of two quantities
\begin{equation}\label{1.3}
S^2\,=\,\frac{\int d^2 b \mid M^H(s,b)\mid^2 P^S(s,b)}
{\int d^2 b \mid M^H(s,b)\mid^2}.
\end{equation}
$M^H(s,b)$ is the amplitude for the LRG 
diffractive process (soft or hard) of
interest. $P^S(s,b)$
is the probability that no inelastic
soft interaction in the re-scattering eikonal chain  
results in inelasticity of the final state 
at $(s,b)$.

The organization of this paper is as follows: 
In Sec.2 we briefly review 
the role of s-channel unitarity in high energy soft scattering 
and the eikonal model.
The GLM model\cite{Gotsman:1992ui,Gotsman:1993ux,Gotsman:1995gg,Gotsman:1994gc} and its consequent survival 
probabilities\cite{Gotsman:1993vd,Gotsman:1996ix,Gotsman:1998mm} are presented in 
Sec.3, including a generalization to a multi channel
re-scattering model\cite{Gotsman:1999ri,Gotsman:1999xq}. 
The KKMR model\cite{Khoze:2000cy,Khoze:2000wk,Khoze:2000jm,Kaidalov:2001iz,Kaidalov:2003ys} and its survival probabilities 
is presented in Sec.4.  
A discussion and our conclusions are presented in Sec.5. 
An added short presentation on Monte Carlo calculations of $S^2$ is 
given in an Appendix.
\section{Unitarity} 

Even though soft high energy scattering has been extensively studied 
experimentally over the last 50 years, we do not have, as yet, a 
satisfactory 
QCD framework to calculate even the gross features of this 
impressive data base. This 
is just a reflection of our inability to execute QCD 
calculations in the non-perturbative regime. 
High energy soft scattering is, thus, commonly described by 
the Regge-pole model\cite{Collins:1977jy,Caneschi}. The theory, motivated by S matrix 
approach, was introduced more than
40 years ago and was soon after followed by a very rich 
phenomenology.

The key ingredient of the Regge pole model  
is the leading Pomeron, whose 
linear $t$-dependent trajectory is given by
\begin{equation}\label{2.1}
\alpha_{\pom}(t)\,=\,\alpha_{\pom}(0)\,+\,\alpha_{\pom}^{\prime} t.
\end{equation}
A knowledge of $\alpha_{\pom}(t)$ enables a calculation of 
$\sigma_{tot},\,\sigma_{el}$ and $\frac{d\sigma_{el}}{dt}$, 
whose forward elastic exponential slope is given by
\begin{equation}\label{2.4.1}
B_{el}\,=\,2 B_0\,+\,2\alpha_{\pom}^{\prime}ln\left(\frac{s}{s_0}\right).
\end{equation}
Donnachie and Landshoff (DL) have vigorously promoted\cite{Donnachie:1992ny,Donnachie:1993it} an 
appealing and very simple Regge parametrization for total and 
forward differential elastic hadron-hadron cross sections in which 
they offer a global fit to all available hadron-hadron and photon-hadron
total and elastic cross section data.
This data, above $P_L=10\,GeV$, 
is excellently fitted with universal parameters.
We shall be interested only in the DL  
Pomeron with an intercept 
$\alpha_{\pom}(0)\,=\,1\,+\,\epsilon$, 
where $\epsilon\,=\,0.0808$, 
which accounts for 
the high energy growing cross sections. Its fitted\cite{Block:1991nd} slope 
value is $\alpha_{\pom}^{\prime}\,=\,0.25\,GeV^{-2}$. 
\subsection{S-channel unitarity} 
The simple DL parametrization is bound to 
violate s-channel unitarity at some energy 
since $\sigma_{el}$ grows with energy 
as $s^{2\epsilon}$, modulu logarithmic corrections, while $\sigma_{tot}$
grows only as $s^{\epsilon}$. 
The theoretical problems at stake are easily identified 
in an impact b-space representation. 

The elastic scattering amplitude is normalized so that
\begin{equation}\label{2.6}
\frac{d\sigma_{el}}{dt}\,=\,\pi\mid f_{el}(s,t) \mid ^2,
\end{equation}
\begin{equation}\label{2.7}
\sigma_{tot}\,=\,4 \pi Im f_{el}(s,0).
\end{equation}
The elastic amplitude in b-space is defined as
\begin{equation}\label{2.8}
a_{el}(s,b)\,=\,\frac{1}{2\pi}\int d{\bf q} e^{-i{\bf q\cdot b}} 
f_{el}(s,t),
\end{equation}
where $t\,=\,-{\bf q}^2$.
In this representation
\begin{equation}\label{2.9}
\sigma_{tot}\,=\,2\int d^2 b\, Im [a_{el}(s,b)],
\end{equation}
\begin{equation}\label{2.10}
\sigma_{el}\,=\,\int d^2 b \mid a_{el}(s,b) \mid ^2,
\end{equation}
\begin{equation}\label{2.10.1}
\sigma_{in}\,=\,\sigma_{tot}\,-\,\sigma_{el}.
\end{equation}
\begin{figure}
\centerline{\epsfxsize=7cm \epsfbox{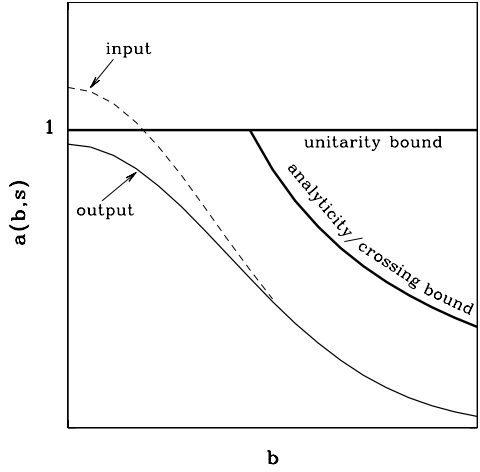}}
\caption{
\it A pictorial illustration of a high energy
b-space elastic amplitude bounded by unitarity and analyticity/crossing.
In the illustration we have an input amplitude which violates the eikonal 
unitarity bound and an output amplitude obtained after a unitarization
procedure.
\label{asb.eps}}
\end{figure}

As noted, a simple Regge pole with $\alpha_{\pom}(0)>1$
will eventually violate s-channel unitarity. 
The question is if this is a future problem to be confronted only at far 
higher energies than presently available, or is it a phenomena which can 
be identified through experimental signatures observed within the 
available high energy data base. 
It is an easy exercise to check that the DL 
model\cite{Donnachie:1992ny,Donnachie:1993it}, with its fitted global parameters, will violate the 
unitarity black bound (see {\bf{2.2}}) 
at very small b, 
just above the present Tevatron energy. 
Indeed, CDF reports\cite{Abe:1993wu} that 
$a_{el}(b=0,\sqrt{s}=1800)=0.96\pm 0.04$. 
A pictorial illustration of the above is presented in Fig.1.
Note that 
the energy dependence of the experimental SD cross section\cite{Goulianos:2003ik,Goulianos:1999wi,Goulianos:2000gp,Goulianos:2004as,Dino}  
in the ISR-Tevatron energy range is
much weaker than the power dependences observed for 
$\sigma_{el}$. 
Diffractive cross sections are not discussed in the DL model. 
\subsection{The eikonal model}
The theoretical difficulties, pointed out in the previous subsection,  
are eliminated once we take into account the corrections 
necessitated by unitarity. The problem is that enforcing unitarity 
is a model dependent procedure. 
In the following we shall confine ourselves  
to a Glauber type eikonal model\cite{Chou:1968bc}. 
In this approximation, the scattering matrix is 
diagonal and only repeated elastic 
re-scatterings are summed. Accordingly, we write
\begin{equation}\label{2.12}
a_{el}(s,b)\,=\,i\left(1-e^{-\Omega(s,b)/2}\right).
\end{equation}   
Since the scattering matrix is diagonal, the unitarity constraint is 
written as
\begin{equation}\label{2.13}
2 Im [a_{el}(s,b)]\,=\,{\mid a_{el}(s,b) \mid}^2\,+\,G^{in}(s,b),
\end{equation}
with
\begin{equation}\label{2.14}
G^{in}\,=\,1\,-\,e^{-\Omega(s,b)}.
\end{equation}
The eikonal expressions for the soft cross sections of interest are 
\begin{equation}\label{2.12.1}
\sigma_{tot}\,=\,2 \int d^2 b
\left(1\,-\,e^{-\Omega(s,b)/2}\right),
\end{equation}
\begin{equation}\label{2.12.2}
\sigma_{el}\,=\,\int d^2 b \left(1\,-\,e^{-\Omega(s,b)/2}\right)^2,
\end{equation}
\begin{equation}\label{2.12.3}
\sigma_{in}\,=\,\int d^2 b \left(1\,-\,e^{-\Omega(s,b)}\right),
\end{equation}
and
\begin {equation}\label{Bel}
B_{el}(s)\,=
\,\frac{\int\,d^2b\,\,b^2\,\left(1\,-\,e^{-\Omega(s,b)/2}\right)}
{2\,\int\,d^2b\,\left(1\,-\,e^{-\Omega(s,b)/2}\right)}.
\end{equation}
From Eq.(\ref{2.14}) it follows that 
$P^S(s,b)\,=\,e^{-\Omega(s,b)}$ is  
the probability that the final state of 
the two initial interacting hadrons is elastic, regardless of the 
eikonal rescattering chain. 
It is identified, thus, with $P^S(s,b)$ of Eq.(\ref{1.3}).

Following our implicit assumption that, in the high energy limit, hadrons 
are correct degrees of freedom, i.e. they diagonalize the interaction 
matrix,
Eq.(\ref{2.12}) is a general solution of Eq.(\ref{2.13}) as
long as the input opacity $\Omega$ is arbitrary. 
In the eikonal model
$\Omega$ is real and equals the imaginary part of
the iterated input Born amplitude.
The eikonalized amplitude is imaginary. Its analyticity
and crossing symmetry are easily restored. 
In a Regge language we substitute, to this end, 
$s^{\alpha_{\pom}}\,\rightarrow
\,s^{\alpha_{\pom}}e^{-\frac{1}{2}i\pi \alpha_{\pom}}$.

In the general case,
Eq.(\ref{2.13}) implies a general bound,  
$\mid a_{el}(s,b)\mid \leq 2$, obtained when $G^{in}=0$. This is an 
extreme option in which asymptotically 
$\sigma_{tot}=\sigma_{el}$\cite{Troshin:2004gh}. This is formally acceptable but 
not very appealing.
Assuming that $a_{el}$ is imaginary, we obtain that 
the unitarity bound coincides with the black disc bound, 
$\mid a_{el}(s,b)\mid \leq 1$. Accordingly, 
\begin{equation}\label{bound}
\frac{\sigma_{el}}{\sigma_{tot}} \leq \frac{1}{2}. 
\end{equation}

\section{The GLM Model}

The GLM screening correction (SC) model\cite{Gotsman:1992ui,Gotsman:1993ux,Gotsman:1995gg,Gotsman:1994gc} 
is an eikonal model originally conceived so
as to explain the exceptionally mild energy dependence of soft diffractive  
cross sections.
It utilized the observation that  
s-channel unitarization enforced by the eikonal
model operates on a diffractive amplitude in a different way than it
does on the elastic amplitude. The GLM 
diffractive damping
factor is identical to Bjorken's survival probability.
\subsection{The GLM SC model}
In the GLM model, we take a DL type Pomeron
exchange amplitude input in which
$\alpha_{\pom}(0)\,=\,1+\Delta\,>0$.
The simplicity of the GLM SC model 
derives from the observation that
the eikonal approximation with a 
central Gaussian input, corresponding to an exponential slope of
$\frac{d\sigma_{el}}{dt}$, can be summed analytically. 
This is, clearly, an over
simplification, but it reproduces the bulk of the data well,
i.e. the total and the forward elastic cross sections.
Accordingly,
the eikonal DL type b-space expression for $\Omega(s,b)$
is: 
\begin{equation}\label{2.15}
\Omega(s,b)\,=\,\nu(s)\,\Gamma^S(s,b),
\end{equation}
where,
\begin{equation}\label{2.16}
\nu(s)\,=\,\sigma(s_0)\,\left(\frac{s}{s_0}\right)^{\Delta},
\end{equation}
\begin{equation}\label{2.17}
R^2(s)\,=\,4R_0^2\,+\,4\alpha_{\pom}^{\prime} ln (\frac{s}{s_0}),
\end{equation}
and the soft profile is defined
\begin{equation}\label{2.18.0}
\Gamma^S(s,b)\,=\,\frac{1}{\pi R^2(s)}\,e^{-\frac{b^2}{R^2(s)}}.
\end{equation}
It is defined so as to keep the normalization 
$\int d^2b \,\Gamma^S(s,b)\,=\,1$.

One has to distinguish between
the eikonal model input and output. The key element is that
the power $\Delta$, and $\nu$, are input information,
not bounded by unitarity, and
should not be confused with DL effective power $\epsilon$ and the
corresponding total cross section.
Since the DL model reproduces the forward
elastic amplitude, in the ISR-HERA-Tevatron range, well, we
require that the eikonal model output will be compatible with
the DL results.
Obviously, $\Delta\,>\,\epsilon$.
In a non screened DL type model with a Gaussian profile the relation
$B_{el}\,=\,\frac{1}{2}R^2(s)$ is exact. In a screened model, like GLM,
$B_{el}\,>\,\frac{1}{2}R^2(s)$ due to screening. 

With this input we get
\begin{equation}\label{2.19.1}
\sigma_{tot}\, =\, 2 \pi R^2(s)
\left[ln\left(\frac{\nu(s)}{2}\right)+C-Ei\left(-\frac{\nu(s)}{2}\right)\right]
\,\propto\, ln^2(s), 
\end{equation}
\begin{equation}\label{2.19.2}
\sigma_{el}\,=\,\pi R^2(s)
\left[ln\left(\frac{\nu(s)}{4}\right)+C-2Ei\left(-\frac{\nu(s)}{2}\right)
+Ei\left(-\nu(s)\right)\right]
\propto\,\frac{1}{2} ln^2(s),
\end{equation}
\begin{equation}\label{2.19.3}
\sigma_{in}\,=\,\pi R^2(s)
\{ln[\nu(s)]+C-Ei[-\nu(s)]\}
\,\propto\, \frac{1}{2} ln^2(s).
\end{equation}
$Ei(x)\,=\,\int_{-\infty}^x \frac{e^t}{t} dt$, and $C\,=\,0.5773$
is the Euler constant.
An important consequence of the above is that the ratio 
$\frac{\sigma_{el}}{\sigma_{tot}}$ is a single variable function 
of $\nu(s)$. In practice it means that given the experimental value of 
this ratio at a given energy we can obtain an "experimental" value of 
$\nu$ which does not depend on the adjustment of free parameters.

The formalism presented above is extended to diffractive channels through  
the observation, traced to Eqs.$(\ref{1.3})$ and $(\ref{2.14})$,
that $P^S(s,b)\,=\,e^{-\Omega(s,b)}$.
Accordingly, a screened non elastic diffractive cross section is obtained
by convoluting its b-space amplitude square with the probability $P^S$.   

The above has been utilized\cite{Gotsman:1992ui,Gotsman:1993ux,Gotsman:1995gg,Gotsman:1994gc} to calculate the soft integrated 
single diffraction cross section. To this end, we write, 
in the triple Regge approximation\cite{Mueller:1970fa}, the double differential 
cross section 
$\frac{M^2d\sigma_{sd}}{dM^2dt}$, where $M$ is the diffracted mass. 
We, then, transform it to b-space, 
multiply by $P^S(s,b)$ and integrate. The output 
$\frac{M^2d\sigma_{sd}}{dM^2dt}$, changes its high energy 
behaviour from $s^{2\Delta}$ modulu $ln (\frac{s}{s_0})$ (which is 
identical  
to the behaviour of a DL elastic cross section) to the moderate behaviour 
of $ln(\frac{s}{s_0})$. 
Note also a major difference in the diffractive b-space profile which 
changes from an input central Gaussian to an output peripheral 
distribution peaking at 
higher b. Consequently, the GLM model 
is compatible with the Pumplin bound\cite{Pumplin:1973cw,Pumplin:1982na}.
\begin{equation}\label{4.15}
\frac{\sigma_{el}(s,b)\,+\,\sigma_{diff}(s,b)}{\sigma_{tot}(s,b)}\,
\leq\,\frac{1}{2}.
\end{equation}
\subsection{Extension to a multi channel model}
The most serious deficiency of a single channel eikonal model is
inherent, as the model considers only elastic rescatterings.
This is incompatible with the relatively large diffractive cross section
observed in the ISR-Tevatron energy range.
To this we add a specific problematic feature of the GLM model.
Whereas, $\sigma_{tot}$, $\sigma_{el}$ and $B_{el}$ are very well
fitted, the reproduction of
$\sigma_{sd}$, in the available ISR-Tevatron range, is poorer.
A possible remedy to these
deficiencies is to replace the one channel
with a multi channel eikonal model,
in which inelastic diffractive intermediate re-scatterings
are included as well\cite{Gotsman:1999ri,Gotsman:1999xq,Gutman}.
However, we have to insure that a
multi channel model does improve the diffractive (specifically
SD) predictions of the GLM model,
while maintaining, simultaneously, its
excellent reproductions\cite{Gotsman:1992ui,Gotsman:1993ux,Gotsman:1995gg,Gotsman:1994gc} of
the forward elastic amplitude,
as well as its appealing results on
LRG survival probabilities\cite{Gotsman:1993vd,Gotsman:1996ix,Gotsman:1998mm} to be
discussed in {\bf{3.3}}.

In the simplest approximation we consider diffraction as a single hadronic
state. We have, thus, two orthogonal wave functions
\begin{equation}\label{4.1}
\langle\Psi_{h} \mid \Psi_{d}\rangle\,=\,0.
\end{equation}
$\Psi_{h}$ is the wave function of the incoming hadron, and
$\Psi_{d}$ is the wave function of the outgoing diffractive system
initiated by the incoming hadron.
Denote the interaction operator by {\bf T} and consider two wave functions
$\Psi_1$ and $\Psi_2$ which are diagonal with respect to {\bf T}.
The amplitude of the interaction is given by
\begin{equation}\label{4.2}
A_{i,k}=
\langle\Psi_i \Psi_k \mid {\bf T} \mid \Psi_{i^{\prime}}
\Psi_{k^{\prime}}\rangle
=a_{i,k}\,\delta_{i,i^{\prime}}\,\delta_{k,k^{\prime}}.
\end{equation}
In a $2 \times 2$ model $i,k\,=\,1,2$. The amplitude $a_{i,k}$
satisfies the diagonal unitarity condition (see Eq.(\ref{2.13}))
\begin{equation}\label{4.3}
2 Im\,a_{i,k}(s,b) \,=\, \mid a_{i,k}(s,b) \mid^2 \,+\,
G_{i,k}^{in}(s,b),
\end{equation}
for which we write the solution
\begin{equation}\label{4.4}
a_{i,k}(s,b) \,=\,
i\left(1\,-\,e^{-\frac{\Omega_{i,k}(s,b)}{2}}\right),
\end{equation}
and
\begin{equation}\label{4.5}
G_{i,k}^{in} \,=\, 1-e^{-\Omega_{i,k}(s,b)}.
\end{equation}
$\Omega_{i,k}(s,b)$ is the opacity of the $(i,k)$ channel
with a wave function $\Psi_i\, \times \,\Psi_k$.
\begin{equation}\label{4.13}
\Omega_{i,k}\,=\,\nu_{i,k}(s)\,\Gamma_{i,k}^S(s,b) 
\end{equation}
where
\begin{equation}\label{4.13.1}
\nu_{i,k}\,=\,\sigma_{i,k}^{S0}\,\(\frac{s}{s_0}\)^\Delta.
\end{equation}
The factorizable radii are given by
\begin{equation}\label{4.14}
R_{i,k}^2(s)\,=\,2R_{i,0}^2\,+\,2R_{0,k}^2\,  
+\,4\alpha_{\pom}^{\prime} ln (\frac{s}{s_0}).
\end{equation}
$\Gamma_{i,k}^S(s,b)$ is the soft profile of the (i,k) channel.
The probability that the final state of two interacting hadron states, 
with quantum numbers i and k, will be elastic regardless of the 
intermediate
rescatterings is
\begin{equation}\label{4.5.1}
P_{i,k}^S(s,b)\,=\,e^{-\Omega_{i,k}(s,b)}\,=\,\{1\,-\,a_{i,k}(s,b)\}^2.
\end{equation}

In the above diagonal representation, 
$\Psi_h$ and $\Psi_d$ can be written as
\begin{equation}\label{4.6}
\Psi_h\,=\,\alpha \Psi_1\,+\,\beta \Psi_2,
\end{equation}
\begin{equation}\label{4.7}
\Psi_d\,=\,-\beta \Psi_1\,+\,\alpha \Psi_2.
\end{equation}
$\Psi_1$ and $\Psi_2$ are orthogonal.
Since $\mid \Psi_h \mid^2\,=\, 1$, we have
\begin{equation}\label{4.8}
\alpha^2\,+\,\beta^2\,=\,1.
\end{equation}
The wave function of the final state is
\begin{eqnarray}\label{4.9}
\lefteqn{\Psi_f\,=\,\mid {\bf T} \mid \Psi_h \times \Psi_h\rangle\,=}
\nonumber \\ & &
\,\alpha^2 a_{1,1} \{\Psi_1 \times \Psi_1\}\,+\,
\alpha \beta a_{1,2} \{\Psi_1 \times \Psi_2\,+\,
\Psi_2 \times \Psi_1 \}\,+\,
\nonumber \\ & &
\beta^2 a_{2,2} \{\Psi_2 \times \Psi_2\}.
\end{eqnarray}
We have to consider 4
possible re-scattering processes. However, in the case of a $\bar p p$
(or $p p$)
collision, single diffraction at the proton vertex equals single
diffraction at the antiproton vertex. i.e., $a_{1,2}\,=\,a_{2,1}$ and we
end with three channels whose
b-space amplitudes are given by
\begin{equation}\label{4.10}
a_{el}(s,b)\,=\, \langle\Psi_h \times \Psi_h \mid
\Psi_f\rangle \,=\, \alpha^4 a_{1,1} \,+\,
2 \alpha^2 \beta^2 a_{1,2}\,+\,
\beta^4 a_{2,2},
\end{equation}
\begin{equation}\label{4.11}
a_{sd}(s,b) \,=\, \langle\Psi_h \times \Psi_d \mid
\Psi_f\rangle \,=\,
\alpha \beta \{\- \alpha^2 a_{1,1}+(\alpha^2-\beta^2) a_{1,2}+
\beta^2 a_{2,2} \},
\end{equation}
\begin{equation}\label{4.12}
a_{dd}(s,b) \,=\, \langle\Psi_d \times \Psi_d \mid
\Psi_f\rangle \,=\,
\alpha^2 \beta^2 \{ a_{1,1}\,-\, 2 a_{1,2}\,+\, a_{2,2}\}.
\end{equation}
In the numeric calculations one may further neglect the
double diffraction channel which is exceedingly small 
in the ISR-Tevatron range.
This is obtained by setting 
$a_{2,2}\,=\,2 a_{1,2}\,-\,a_{1,1}$.
Note that in the limit where $\beta\,<<\,1$, 
we reproduce the single channel model.

\begin{figure}
\centerline{ \epsfxsize=7cm \rotatebox{-90}{\epsfbox{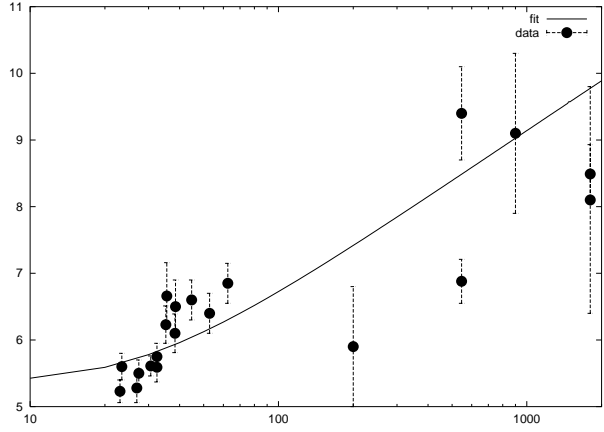}}}
\caption{
{\it Integrated SD data and a two channel model fit.}
\label{zursd}}
\end{figure}

As in the single channel,
we simplify the calculation assuming a Gaussian b-space distribution of
the input opacities soft profiles
\begin{equation}\label{4.5.0}
\Gamma_{i,k}^S(s,b)\,=
\,\frac{1}{\pi R_{i,k}^2(s)}\,e^{-\frac{b^2}{R_{i,k}^2(s)}}.
\end{equation}
The opacity expressions, just presented, allow
us to express the
physical observables of interest as
functions of $\nu_{1,1},\,\nu_{1,2},\,R_{1,1}^2,\,R_{1,2}^2$  
and $\beta$, which is a constant of the model.
The determination of these variables
enables us to produce a global
fit to the total, elastic and diffractive 
cross sections as well as
the elastic forward slope.
This has been done in a two channel
model, in which $\sigma_{dd}$ is neglected\cite{Gotsman:1999ri}. 
The main conclusion of this study is that the
extension of the GLM model to a multi channel eikonal
results with a very good
overall reproduction of the data. 
The results maintain the
b-space peripherality of the diffractive output amplitudes and
satisfy the Pumplin bound\cite{Pumplin:1973cw,Pumplin:1982na}.
Note that since different experimental groups
have been using different algorithms to define diffraction, the SD
experimental points are too scattered to enable a tight
theoretical reproduction of the diffractive data, see Fig.\ref{zursd}.
\subsection{Survival probabilities of LRG in the GLM model}
The eikonal model simplifies the calculation of the survival 
probability, Eq.({\ref{1.3}), 
associated with the soft re-scatterings of
the spectator partons. 
We can, thus, eliminate the 
nominator and denominator terms in 
\newline
$\mid M^H(s,b)\mid^2$ 
which depend exclusively on s.
In the GLM model we assume a Gaussian b-dependence for
$\mid M^H(s,b) \mid^2$ 
corresponding to a constant hard radius
${R^H}^2$. This choice
enables an analytic solution of Eq.(\ref{1.3}). More elaborate choices, 
such as dipole or multi poles distributions, 
require a numerical evaluation of this equation.

Define,
\begin{equation}\label{3.3.1}
a_H(s)\,=\,\frac{R^2(s)}{{R^H}^2(s)}>1.
\end{equation}
$a_H(s)$ grows logarithmically with $s$.
As stated, Eq.(\ref{1.3}) can be analytically evaluated with our choice 
of Gaussian profiles and we get
\begin{equation}\label{3.4}
S^2\,=\,\frac{a_H(s) \gamma[a_H(s),\nu(s)]}
{[\nu(s)]^{a_H(s)}},
\end{equation}
where 
$\gamma(a,\nu)$ denotes the incomplete Euler gamma function
\begin{equation}\label{3.4.1}
\gamma(a,x)\,=\,\int_0^x\,z^{a-1}e^{-z}dz.
\end{equation}

The solution of Eq.(\ref{3.4}), at a given $s$, depends on the 
input values of ${R^H}^2$, $R^2$ and $\nu(s)$. 
In the GLM approach, ${R^H}^2$ is estimated from the excellent
HERA data\cite{Adloff:2000vm,Chekanov:2002xi,Kowalski:2003hm} on $\gamma+p \rightarrow J/\Psi+p$.
The values of $\nu(s)$ and $R^2(s)$ are obtained from the experimental 
$\bar p p$ data. This can be attained from a global fit to the soft 
scattering data\cite{Gotsman:1999ri}. 
Alternatively, we can obtain $\nu$ from the ratio 
$\frac{\sigma_{el}}{\sigma_{tot}}$ and then obtain the value of $R^2$ 
from the explicit expressions given in 
Eqs.(\ref{2.19.1},\ref{2.19.2},\ref{2.19.3}).
LHC predictions presently depend on 
model calculations with which this
information can be obtained.
Once we have determined $\nu(s)$ and $a_H(s)$, 
the survival probability is
calculated from Eq.(\ref{3.4}). 

In the GLM three channel model we obtain for central 
hard diffraction of di-jets or Higgs a survival probability, 
\begin{equation}\label{3.7.1}
S_{CD}^2(s)\,=\,\frac{\int d^{2}b\,
\left(\alpha^4\,P_{1,1}^S\,{\Omega_{1,1}^H}^2\,+
\,2 \alpha^2\beta^2\,P_{1,2}^S\,{\Omega_{1,2}^H}^2\,+
\,\beta^4 P_{2,2}^S\,{\Omega_{2,2}^H}^2\right)}
{\int\,d^{2}b
\left(\alpha^4\,{\Omega_{1,1}^H}^2\,+
\,2 \alpha^2 \beta^2\,{\Omega_{1,2}^H}^2\,+
\,\beta^4\,{\Omega_{2,2}^H}^2\right)}.
\end{equation}

The hard diffractive cross sections 
in the (i,k) channel are calculated 
using the multi particle optical theorem\cite{Mueller:1970fa}. 
They are written in the same form as the soft amplitudes 
\begin{equation}\label{3.6}
{\Omega_{i,k}^H}^2\,=\,{\nu_{i,k}^H(s)}^2\,\Gamma_{i,k}^H(b),
\end{equation}
where,
\begin{equation}\label{3.6.1}
\nu_{i,k}^H\,=\,\sigma_{i,k}^{H0}\(\frac{s}{s_0}\)^{\Delta_H}.
\end{equation}
As in the single channel calculation 
we assume that $\Gamma_{i,k}^H(b)$ is Gaussian, 
\begin{equation}\label{3.6.2}
\Gamma_{i,k}^H(b)\,=
\,\frac{2}{\pi R_{i,k}^2}\,e^{-\frac{2\,b^2}{R_{i,k}^2}}.
\end{equation}
Note, that the hard radii ${R_{i,k}^H}^2$ 
are constants derived from HERA $J/\Psi$ photo 
and DIS production\cite{Adloff:2000vm,Chekanov:2002xi,Kowalski:2003hm}. 

As it stands, a three channel calculation is not useful since
$\sigma_{dd}$ is very small and the 3'd channel 
introduces additional parameters which can not be constraint  
by the meager experimental information on $\sigma_{dd}$\cite{Goulianos:2003ik,Goulianos:1999wi,Goulianos:2000gp,Goulianos:2004as,Dino}.
In a two channel model Eq.(\ref{3.7.1}) reduces to
\begin{equation}\label{3.7.2}
S_{CD}^2(s)\,=\,\frac{\int d^{2}b\,
\left(\,P_{1,1}^S\,{\Omega_{1,1}^H}^2\,-
\,2 \beta^2\,(P_{1,1}^S\,{\Omega_{1,1}^H}^2\, 
-\,P_{1,2}^S\,{\Omega_{1,2}^H}^2)\right)}
{\int\,d^{2}b
\left(\,{\Omega_{1,1}^H}^2\,-
\,2 \beta^2\,({\Omega_{1,1}^H}^2\,-\,{\Omega_{1,2}^H}^2)\right)}.
\end{equation}
A new, unpublished yet, model\cite{Gotsman}, offers
an explicit $S^2$ calculation for the exclusive 
$NN \rightarrow N+LRG+2J+LRG+N$ final state,
both in one and two channel eikonal models.
We shall comment on its output in the next subsection.

\subsection{GLM $S^2$ predictions}
\label{sec:S2pred}
Following are a few general comments on the GLM calculations of $S^2$, 
after which we discuss the input/output features of 
the single and two channel models.  
Our objective is to present predictions for LHC.

The only available experimental observable 
with which we can check the theoretical $S^2$ predictions
is the hard LRG di-jets data 
obtained in the Tevatron and Hera.
A comparison between data and our predictions is not immediate as the
basic measured observable is $f_{gap}$ and not $S^2$.
The application of the GLM models to a calculation of $f_{gap}$
depends on an external input of a hard diffractive LRG
cross section which is then corrected by $S^2$ as
presented above. Regardless of this deficiency, the introduction of a
survival probability is essential so as to understand the huge difference
between the pQCD calculated $F_{gap}$ and its experimental value 
$f_{gap}$.
A direct test of the GLM predictions calls for 
a dedicated experimental determination of $S^2$.
The only direct $S^2$ information from the Tevatron  
is provided by a JGJ ratio measured by
D0\cite{Abachi:1994hb,Abachi:1995gz,Abbott:1998jb} in which
$\frac{S^2({\sqrt{s}\,=\,630})}
{S^2({\sqrt{s}\,=\,1800})}\,=\,2.2 \pm 0.8$.
This is to be compared with a GLM ratio of 
$1.2-1.3 \pm 0.4$ presented below. 

The survival probabilities of the CD, SD and
DD channels are not identical. The key difference is that
each of the above channels has a different hard radius.
A measure of the sensitivity
of $S^2$ to changes in $\nu$ and $a_H$ is easy to identify 
in a single channel calculation which is presented in Fig.3.
\begin{figure}
\begin{center}
\includegraphics[width=7cm,bb=90 0 380 290]{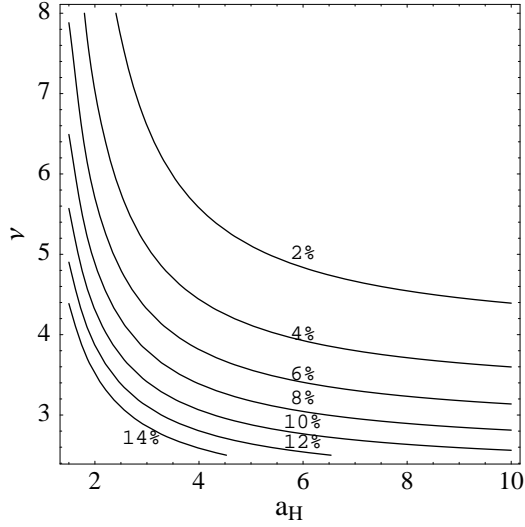} 
\end{center}
\caption{\it A contour plot of $S^2(1C)$ against $\nu(s)$ and $a_H(s)$.}
\label{nua}
\end{figure}
Indeed, preliminary CDF GJJG data\cite{Dino} suggest that $f_{gap}$
measured for this channel is moderately smaller than the 
rate measured for the GJJ channel.

GLM soft profile input is a
central Gaussian. This is over simplified, and most models
assume a power like dipole or multipole b-dependence of $\Gamma^S(s,b)$
and $\Gamma^H(s,b)$.
Explicit comparisons\cite{Gotsman}
of $S^2$ obtained with different input profiles
shows a diminishing difference between the 
survival probability outputs, provided
their effective radii are compatible.

Regardless of the attractive simplicity of the single channel model,
one should add a cautious reminder 
that the single channel model does not reproduce
$\sigma_{sd}$ well since its survival probabilities are over-estimated. 
Consequently, we are 
inclined to suspect that the $S^2$
values presented in the table below are over-estimated as well.
\begin{center}
\begin{tabular}{|c|c|c|c|c|c|c|}
\hline
\multicolumn{1}{|c}{$\sqrt{s}\,\,$(GeV)} &
\multicolumn{1}{|c}{$S_{CD}^2(F1C)$} &
\multicolumn{1}{|c}{$S_{CD}^2(D1C)$} &
\multicolumn{1}{|c}{$S_{SD_{incl}}^2(F1C)$} &
\multicolumn{1}{|c}{$S_{SD_{incl}}^2(D1C)$} &
\multicolumn{1}{|c}{$S_{DD}^2(F1C)$} &
\multicolumn{1}{|c|}{$S_{DD}^2(D1C)$} \\
\hline
  540 & 14.4\% &13.1\% & 18.5\% &17.5\% &22.6\% & 22.0\% \\
 1800 & 10.9\% & 8.9\% & 14.5\% &12.6\% &18.2\% & 16.6\% \\
14000 &  6.0\% & 5.2\% &  8.6\% & 8.1\% &11.5\% & 11.2 \% \\
\hline
\end{tabular}
\end{center}

As we noted, the soft input can be obtained from either 
a model fit to the soft scattering data or directly 
from the measured values 
of $\sigma_{tot}, \sigma_{el}$ and ${R^H}^2$. The first method is denoted 
F1C and the second is denoted D1C. Note that having no LHC data,  
$S_{DD}^2(D1C)$, at this energy, is
calculated on the basis of model estimates for the total and elastic
cross sections. The constant hard radius 
${R^H}^2\,=\,7.2$ is deduced from HERA $J/\Psi$ photoproduction 
forward exponential slope which shows only diminishing 
shrinkage\cite{Adloff:2000vm,Chekanov:2002xi}.
This is a conservative choice which may be  changed slightly with the
improvement of the Tevatron CDF estimates\cite{Abe:1997bp}
of ${R^H}^2$.
The two sets of results obtained 
are compatible, even though, $S^2(D1C)$ is consistently 
lower than $S^2(F1C)$. 
The $S^2$ output presented above depends
crucially  on the
quality of the data base from which we obtain the
input parameters. The two sets of Tevatron data
at $1800\,GeV$ have a severe $10-15\%$ difference resulting in  
a non trivial ambiguity of the $S^2$ output.

The global GLM two channel fit\cite{Gotsman:1999ri} reproduces 
the soft scattering data (including SD) remarkably well with 
$\beta\,=\,0.464$. Its fitted parameters are used for the soft input 
required for the $S^2$ calculations.
Our cross section predictions for LHC 
are: $\sigma_{tot}=103.8\,mb$, $\sigma_{el}=24.5\,mb$, 
$\sigma_{sd}=12\,mb$
and $B_{el}=20.5\,GeV^{-2}$. 
The input for the calculation of $S^2$
requires, in addition to the soft parameters, also the values of
${\nu_{i,k}^H}$ and ${R_{i,k}^H}^2$. The needed hard radii can be 
estimated, at present, only for the CD channel, where we
associate the hard radii $R_{1,1}^H$
with the hard radius obtained in
HERA exclusive $J/\Psi$ photoproduction\cite{Adloff:2000vm,Chekanov:2002xi} and
$R_{1,2}^H$ with HERA inclusive $J/\Psi$ DIS production\cite{Kowalski:2003hm}.
Accordingly, we have
${R_{1,1}^{H}}^2=7.2\,GeV^{-2}$, and
${R_{1,2}^{H}}^2=2.0\,GeV^{-2}$.
We do not have experimental input to determine ${\nu_{i,k}^H}$. 
We overcome this difficalty by assuming a Regge-like factorization 
${\sigma_{i,k}^{H0}}/{\sigma_{i,k}^{S0}}\,=\,constant$.
Our predictions for the CD survival probabilities are: 
$6.6\%$ at $540\,GeV$, $5.5\%$ at $1800\,GeV$ and $3.6\%$ at 
$14000\,GeV$. 

These results may be compared with a recent,
more elaborate, eikonal formulation\cite{Gotsman}
aiming to calculate the survival probability of a final 
exclusive $N+LRG+2J(or H)+LRG+N$ state.
These calculations were done in one and two channel models.
The one channel $S_{CD}^2$ predicted values are $14.9\%$ at $540\,GeV$,
$10.8\%$ at $1800\,GeV$ and $6.0\%$ at $14000\,GeV$. 
These values are remarkably similar
to the GLM one channel output.
In the two channel calculations the 
corresponding predictions are 
$5.1\%$, $4.4\%$ and $2.7\%$, which are marginally 
smaller than the GLM two channel output numbers.

In our assessment, the two channel calculations 
provide a more reliable estimate of $S^2$ since they reproduce well 
the soft scattering forward data. Our estimate for the survival 
probability associated with LHC Higgs production is $2.5\%-4.0\%$.

\section{The KKMR Model}
\renewcommand{\thefootnote}{\arabic{footnote}}
The main part of this section ({\bf 4.1-4.3}) was written by V.A. Khoze, 
A.D. Martin and M. Ryskin (KMR) and is presented here without any 
editing.

The KKMR model calculation\cite{Khoze:2000cy,Khoze:2000wk,Khoze:2000jm,Kaidalov:2001iz,Kaidalov:2003ys} of the survival probabilities 
is conceptually quite similar to the GLM model, 
in as much as unitarization is enforced through an eikonal model whose 
parameters provide a good reproduction of the high energy soft 
scattering data. However, the GLM model is confined to a 
geometrical calculation of $S^2$ for which we  
need just the value of ${R^H}^2$, without any 
specification of the hard dynamics. This value is an external input to the 
model.
The KKMR model contains also a detailed pQCD  
calculation of the hard diffractive proccess, specifically, central 
diffractive Higgs production.
Consequently, it can predict a cross section for the channel under 
investigation.  
\subsection{ KKMR model for soft diffraction}
The KMR description \cite{Khoze:2000wk} of soft diffraction in high energy $pp$
(or $p\bar{p}$) collisions embodies \begin{itemize} \item[(i)]
{\it pion-loop} insertions in the bare Pomeron pole, which represent
the nearest singularity generated by $t$-channel unitarity, \item[(ii)]
a {\it two-channel eikonal} which incorporates the Pomeron cuts
generated by elastic and quasi-elastic (with $N^*$ intermediate states)
$s$-channel unitarity, \item[(iii)] high-mass {\it diffractive
dissociation}.  \end{itemize}

The KKMR model gives a good description of the data on the total and 
differential
elastic cross section
throughout the ISR-Tevatron energy interval, see \cite{Khoze:2000wk}.
Surprisingly, KMR found the
bare Pomeron parameters to be
\begin{equation}
\label{eq:a5}
\Delta \; \equiv \; \alpha (0) \: - \: 1 \: \simeq \: 0.10, \quad\quad
\alpha^\prime \; = \; 0.
\end{equation}
On the other hand it is known that the same data can be described by a
simple effective
Pomeron pole with \cite{Donnachie:1983hf,Donnachie:1992ny,Donnachie:1993it}
\begin{equation}
\label{eq:a6}
\alpha_{\pom}^{\rm eff} (t) \; = \; 1.08 \: + \: 0.25~t.
\end{equation}
In this approach the shrinkage of the diffraction cone comes not from the
bare pole
($\alpha^\prime = 0$), but has components from the three ingredients,
(i)--(iii), of the model.
That is, in the ISR-Tevatron energy range
\begin{equation}
\label{eq:a7}
{\rm \lq\lq} \alpha^\prime_{\rm eff} \mbox{\rq\rq} \; = \; (0.034 \: + \:
0.15 \: + \: 0.066)~{\rm
GeV}^{-2}
\end{equation}
from the $\pi$-loop, $s$-channel eikonalisation and diffractive
dissociation respectively.
Moreover, eikonal rescattering suppresses the growth of the cross section
and so $\Delta
\simeq 0.10 > \Delta_{\rm eff} \simeq 0.08$.

Since the model \cite{Khoze:2000wk} embodies all the main features of soft
diffraction KMR expect it to give reliable predictions for the {\it
survival probability} $S^2$ of the rapidity gaps against population by
secondary hadrons from the underlying event, that is hadrons
originating from soft rescattering.  In particular, KMR predict $S^2 =
0.10 ~(0.06)$ for single diffractive events and $S^2 = 0.05~
(0.03)$  for exclusive Higgs boson production, $pp
\rightarrow p + H + p$, at Tevatron (LHC) energies.
\subsection{Calculation of the exclusive Higgs signal}
The basic mechanism for the exclusive process, $pp\to p+H+p$, is
shown in Fig.~$\ref{fig:H1}$. 
The left-hand gluon $Q$ is needed to screen the
colour flow caused by the active gluons $q_1$ and $q_2$.
\begin{figure}
\begin{center}
\centerline{\epsfxsize=8cm \epsfbox{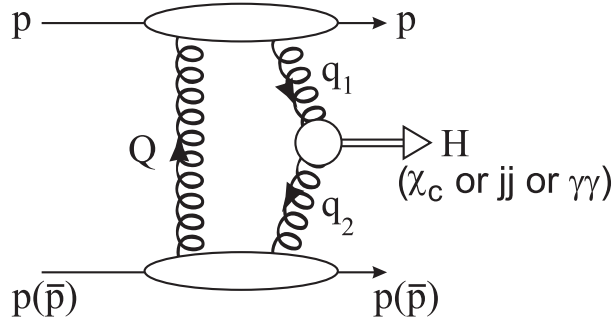}}
\caption{Schematic diagram for central exclusive  production,
$pp \to p+X+p$. The presence of Sudakov form factors ensures the infrared
stability of the $Q_t$ integral over the gluon loop. It is also necessary
to compute the probability, $S^2$, that the rapidity gaps survive soft
rescattering.}
\label{fig:H1}
\end{center}
\end{figure}
Since the dominant contribution comes from the
region $\Lambda_{\rm QCD}^2\ll Q_t^2\ll M_H^2$, the amplitude may
be calculated using perturbative QCD techniques \cite{Khoze:1997dr,Khoze:2000cy}
\begin{equation}
{\cal M}_H \simeq N\int\frac{dQ^2_t}{Q^4_t}\: f_g(x_1, x_1', Q_t^2,
\mu^2)f_g(x_2,x_2',Q_t^2,\mu^2), \label{eq:M} \end{equation} where the
overall normalisation constant $N$ can be written in terms of the $H\to
gg$ decay width \cite{Khoze:2001xm,Khoze:2000cy}.
The probability amplitudes ($f_g$) to find the appropriate pairs of
$t$-channel gluons ($Q,q_1$) and ($Q,q_2$) are given by the skewed
unintegrated gluon densities at the hard scale $\mu$, taken to be
$0.62\  M_H$. Since the momentum fraction $x'$ transfered through the
screening gluon $Q$ is much smaller than that ($x$) transfered through
the active gluons $(x'\sim Q_t/\sqrt s\ll x\sim M_H/\sqrt s\ll 1)$, it
is possible to express $f_g(x,x',Q_t^2,\mu^2)$, to single log accuracy,
in terms of the conventional integrated density
$g(x)$\cite{Kimber:2001sc,Kimber:1999xc,Shuvaev:1999ce,Martin:2001ms}.  The $f_g$'s embody a Sudakov suppression
factor $T$, which ensures that the gluon does not radiate in the
evolution from $Q_t$ up to the hard scale $\mu \sim M_H/2$, and so
preserves the rapidity gaps.

It is often convenient to use the
simplified form \cite{Khoze:2000cy}
\begin{equation}
\label{eq:a61}
  f_g (x, x^\prime, Q_t^2, \mu^2) \; = \; R_g \:
\frac{\partial}{\partial \ln Q_t^2}\left [ \sqrt{T_g (Q_t, \mu)} \: xg
  (x, Q_t^2) \right ], \end{equation} which holds to 10--20\%
accuracy.\footnote{In the actual computations
a more precise form, as given by Eq.~(26) of \cite{Martin:2001ms}, was used.}
The factor $R_g$ accounts for
the single $\log Q^2$ skewed effect \cite{Shuvaev:1999ce}.  It is found to
be about 1.4 at the Tevatron energy and about 1.2 at the energy of the LHC.
\subsection{The Sudakov factor}
The Sudakov factor $T_g (Q_t, \mu)$ reads\cite{Kimber:2001sc,Kimber:1999xc,Watt:2003mx}
\begin{equation}
\label{eq:a71}
T_g (Q_t, \mu)=\exp \left (-\int_{Q_t^2}^{\mu^2}
  \frac{\alpha_S (k_t^2)}{2 \pi}\frac{dk_t^2}{k_t^2} \left[
  \int_\Delta^{1-\Delta}zP_{gg} (z)dz
\ + \ \int_0^1 \sum_q\
  P_{qg} (z)dz\right]\right),
\end{equation}
with $\Delta = k_t/(\mu + k_t)$.  The square root arises in
(\ref{eq:a61}) because the (survival) probability not to emit any
additional gluons  is only relevant to
the hard (active) gluon.  It is the presence of this Sudakov factor
which makes the integration in (\ref{eq:M}) infrared stable, and
perturbative QCD applicable\footnote{Note also that the Sudakov factor 
inside $t$
integration induces an additional strong decrease (roughly as $M^{-3}$
\cite{Kaidalov:2003ys}) of the cross section as the mass $M$ of the centrally
produced hard system increases.  Therefore, the price to pay for
neglecting this suppression effect would be to considerably
overestimate the central exclusive cross section at large masses.}.

It should be emphasized that the presence of the double logarithmic
$T$-factors is a purely classical effect, which was first discussed in
1956 by Sudakov in QED.  There is strong bremsstrahlung when two
colour charged gluons `annihilate' into a heavy neutral object and the
probability not to observe such a bremsstrahlung is given by the
Sudakov form factor\footnote{It is worth mentioning that the $H\to gg$
  width and the normalization factor $N$ in (\ref{eq:M}) is an
  `inclusive' quantity which includes all possible bremsstrahlung
  processes. To be precise, it is the sum of the $H\to gg+ng$ widths,
  with $n$=0,1,2,... . The probability of a `purely exclusive' decay
  into two gluons is nullified by the same Sudakov suppression.}.
Therefore, any model (with perturbative or non-perturbative gluons)
must account for the Sudakov suppression when producing exclusively a
heavy neutral boson via the fusion of two coloured particles.

More details of the role of the Sudakov suppression can be found in J. 
Forshaw's
review in these proceedings\cite{Forshaw:proc}.  Here KMR would like to recall that the
$T$-factors in \cite{Kaidalov:2003fw,Kaidalov:2003ys} were calculated to {\it single} log
accuracy. The collinear single logarithms were summed up using the
DGLAP equation. To account for the `soft' logarithms (corresponding
to the emission of low energy gluons) the one-loop virtual correction
to the $gg\to H$ vertex was calculated explicitly, and then the scale
$\mu=0.62\ M_H$ was chosen in such a way that eq.(\ref{eq:a71})
reproduces the result of this explicit calculation. It is sufficient to
calculate just the one-loop correction since it is known that the
effect of `soft' gluon emission exponentiates. Thus
(\ref{eq:a71}) gives the $T$-factor to single log accuracy.

In some sense, the $T$-factor may be considered as a `survival'
probability not to produce any hard gluons during the $gg\to H$ fusion
subprocess. However, it is not just a number (i.e. a numerical factor) 
which
may be placed in front of the integral (the `bare amplitude'). Without the
$T$-factors hidden in the unintegrated gluon densities $f_g$ the integral
(\ref{eq:M}) diverges. From the formal point of view, the suppression of
the amplitude provided by $T$-factors is infinitely strong, and without 
them
the integral depends crucially on an ad hoc infrared cutoff.
\subsection{Summary of KKMR $S^2$ predictions}
A compilation of $S^2$ values obtained in the KKMR model is presented 
below:
\begin{center}
\begin{tabular}{|c|c|c|c|}
\hline
\multicolumn{1}{|c|}{$\sqrt{s}\,\,$(GeV)} &
\multicolumn{1}{c|}{$S_{2C}^2(CD)$} &
\multicolumn{1}{c|}{$S_{2C}^2(SD_{incl})$} &
\multicolumn{1}{c|}{$S_{2C}^2(DD)$} \\
\hline
  540 & 6.0\% & 13.0\% & 20.0\% \\
 1800 & 4.5\% & 10.0\% & 15.0\% \\
14000 & 2.6\% &  6.0\% & 10.0\% \\
\hline
\end{tabular}
\end{center}
A comparison with the corresponding GLM two channel model is possible only 
for the available GLM CD channel, where,
the KKMR output is compatible with GLM. 
KKMR SD and DD output 
are compatible with the corresponding GLM single channel numbers. 
Overall, we consider the two models to be in a reasonable agreement.

A remarkable utilization of the KKMR model is attained when comparing the 
HERA\cite{Derrick:1993xh,Derrick:1995wv,Derrick:1995tw,Derrick:1995pb,Breitweg:1998gc,Ahmed:1994nw,Ahmed:1995ns,Adloff:1997sc,schilling,Savin:2002eg} and CDF\cite{Abe:1994de,Abe:1997ie,Abe:1998ip,Affolder:2000vb,Affolder:2000hd,Dino} 
di-jets diffractive structure functions derived for the 
dynamically similar GJJ channels.
To this end, the comparison is made between the kinematically 
compatible HERA $F_{jj}^D(Q^2=75\,GeV^2,\beta)$ and the CDF 
$F_{jj}^D(<E_T^2>=75\,GeV^2,\beta)$. 
The theoretical expectation is that $F_{jj}^D(\beta)$, as 
measured by the two experiments, should be very similar. 
\begin{figure}
\begin{center}
\centerline{\epsfxsize=7cm \epsfbox{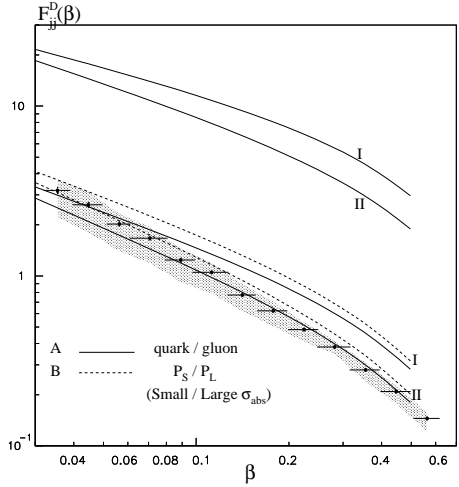}}
\caption{The predictions for the diffractive di-jets production at the
Tevatron (lower lines), obtained from two alternative sets of HERA
diffractive parton distributions I and II, compared with the CDF data
(shaded area). The upper lines correspond to the Tevatron prediction
neglecting the survival probability correction.}
\label{fig:H}
\end{center}
\end{figure}
As can be seen in Fig.\ref{fig:H}, 
the normalizations of the two distributions differ by
approximately an order of magnitude and for very
small $\beta<0.15$ there is a suggestive change in the
CDF distribution shape. This large discrepancy implies 
a breaking of QCD and/or Regge factorization.
Reconsidering, it is 
noted, that HERA DIS data is measured at a high $Q^2$ where
the partonic interactions induced by the highly virtual
photon are point like and, hence, $S^2\,=\,1$.
On the other hand, CDF GJJ measurement is carried out at $1800\,GeV$ 
and, as we saw, its survival probability is rather small.
The convolution between the 
HERA determined GJJ $F_{jj}^D(\beta)$ and 
the $\beta$ dependent survival probabilities, as calculated by KKMR,  
provides the $F_{jj}^D(\beta)$ distribution corrected for the soft 
rescattering of the spectator partons.
This is shown in Fig.\ref{fig:H} and provides an impressive 
reproduction of the experimental distribution. 
We were informed\cite{Khoze} that this analysis 
was successfully redone with an updated 
H1 produced structure function distribution. 

The weak element in the above analysis is that it is crucially dependent 
on the H1 determined $F_{jj}^D(\beta)$ distribution. 
ZEUS has constructed a somewhat different structure function. Clearly, 
a very different experimental determination of $F_{jj}^D(\beta)$, such 
as been recently suggested by Arneodo\cite{Arneodo}, will re-open this 
analysis for further studies, experimental and theoretical. 
\subsection{A Comparison between KKMR and GLM}
The approach of GLM and KKMR to the calculation of forward soft scattering 
in the ISR-Tevatron range are basically similar. Both models utilize the 
eikonal model assuming different input soft profiles which have, nevertheless, compatible effective radii. There are, though, a few particular 
differences between the two sets of calculations:
\newline
1) The GLM model, with a Gaussian soft profile, is applicable only in 
the forward cone ($|t|\,<\,0.3\,GeV^2$), where we have most of the 
data of interest. 
KKMR use a multipole power behaviour profile which enables 
applicability over a, somewhat, wider t range, $|t|\,<\,0.5\,GeV^2$.
Note that, the GLM output is not significantly 
changed with a multipole power behaviour profile provided its radii are 
compatible with the Gaussian input\cite{Gotsman}. 
\newline
2) The GLM input Pomeron trajectory is specified by $\Delta \,=\,0.12$ and 
$\alpha_{\pom}^{\prime}\,=\,0.2$. These evolve 
due to eikonalisation to an effective output 
of $\epsilon\,=\,0.08$ and $\alpha_{\pom}^{\prime}\,=\,0.25$. 
Note that, $\Delta$ is obtained in GLM as a fitted output parameter. 
In KKMR, the relatively high input $\Delta\,\simeq\,0.2$ is theoretically 
tuned by a pion loop renormalization 
resulting in an input value of $\Delta\,\simeq\,0.1$. 
KKMR have a 
more elaborate treatment of $\alpha_{\pom}(t)$ than GLM, 
resulting, nevertheless, with forward cone output predictions similar 
to GLM. However, KKMR accounts for a somewhat wider t range than GLM 
and reproduces the t dependence of $B_{el}$ well. 
Similar results are obtained in a GLM version\cite{Gotsman:1999xq,Gutman} in which the 
soft profile is given by a dipole distribution.
KKMR can predict    
a few differential properties of $S^2$, 
which are beyond the scope of GLM. 
\newline
3) Both models treat the high mass diffraction with the triple Pomeron 
formalism\cite{Mueller:1970fa}. In GLM the final SD cross section is obtained 
by a convolution of the input $\frac{d\sigma_{sd}}{d^2b}$ with 
$P^S(s,b)$. In KKMR the treatment of the SD amplitude is more 
elaborate, ending, though, with no detailed SD data reconstruction which 
is presented in GLM.
\newline
4) The LHC predictions of the 
two models for cross sections and slopes are compatible, with 
the exception of $\sigma_{dd}$ which is neglected in GLM and acquires a 
significant KKMR predicted value of $9.5\,mb$.

GLM is a geometrical model where both the input hard LRG non corrected 
matrix element squared 
and the soft elastic scattering amplitude, are approximated by  
central Gaussians in b-space. 
This property enables us to easily calculate the survival probabilities 
which depend on $\nu$, $R^2$ and ${R^H}^2$ in a single channel input, 
and on $\nu_{i,k}$, $R_{i,k}^2$ and ${R_{i,k}^H}^2$ 
in a two channel input. 
As we have noted, the GLM model, on its own, cannot provide a calculation
of $F_{gap}$ and $f_{gap}$ as it needs the hard radii as an 
external input.
The KKMR model is more sophisticated.
This is attributed to the fact that
the hard diffractive LRG process is 
explicitly calculated in pQCD, hence the non corrected $F_{gap}$ 
and the corrected $f_{gap}$ and $F_{jj}^D$ are model predictions. 
As we have just noted, 
given the hard diffractive matrix element, the actual calculation of the 
diffractive LRG survival probability damping is almost identical to 
GLM. Keeping this basic observation in mind, it is constructive to compare 
the features of the two models with a special interest on the input 
assumptions and output differences of the two models.   

The main difference between the two models is reflected in 
the level of complexity of 
their inputs. GLM soft input is obtained from a simple eikonal model for 
the soft forward scattering, to which we add the 
hard radii which are derived from the HERA data. 
KKMR calculations of $P^S$ are equally simple. The calculation of the 
hard sector matrix elements are, naturally, more cumbersome. 
Given HERA $F_{jj}^D(Q^2,\beta)$, 
a Tevatron diffractive $F_{jj}^D$
in which $<E_T>$ and $Q^2$ are comparable, can be calculated, 
parameter free, without the need 
to calculate the hard amplitude. But this is a particular case and, in 
general, the KKMR calculation depends on 
an extended parameter base, such as the the input p.d.f. and pQCD cuts.
These input parameters are not constrained tightly enough. 

The elaborate structure of the KKMR model provides a rich discovery 
potential which is reflected in the model being able to define 
and calculate the dependence of $S^2$ not only on b, but also on other 
variables, notably $\beta$, and experimental cuts such as the recoil 
proton 
transverse momentum. GLM depends on the hard radii external 
information obtained from HERA data. 
It lacks the potential richness of KKMR. GLM can serve, 
though, as a standard through which we can compare different unitarized 
models. Given such a model, we can extract effective values for 
$\nu$, $R^2$ and ${R^H}^2$ and proceed to a simple calculation of $S^2$.
We shall return to this proposed procedure in the final discussion.

Even though both GLM and KKMR are two channel models, they
are dynamically different. GLM two channel formulation relates to the
diversity of the intermediate soft re-scatterings, i.e. elastic and 
diffractive for which we have different soft amplitudes $a_{i,k}$, each of 
which is convoluted with a different probability $P_{i,k}^S$ 
which depends on a different interaction radius $R_{i,k}^2$.
In the KKMR model the two channels relate to two 
different dynamical options of the hard process. 
In model A the separation is between 
valence and sea interacting partons. 
In model B the separation is between 
small and large dipoles. 
The two models give compatible results. 
The key point, though, is that the KKMR opacities $\Omega_{i,k}$, in the 
definition of $P_{i,k}^S$, differ in their normalization, 
but have the same b-dependence. 
Regardless of this difference the output of the 
GLM and KKMR models is reasonably compatible. 
The compatibility between GLM and KKMR is not surprising since the 
explicit KKMR calculation of the hard LRG amplitude  
is approximated relatively well by the GLM simple Gaussian.

Our final conclusion is that the two model output sets are  
compatible. The richness of the KKMR model has a significant 
discovery potential lacking in GLM. On the other hand, the GLM 
simplicity makes it very suitable as a platform to present different 
models in a uniform way, which enables a transparent comparison.
\section{Discussion}
As we shall see, at the end of this section, there is no significant 
difference between the values of $\sigma_{tot}$ predicted by DL and GLM
up to the top Cosmic Rays energies.
This is, even though, DL is a Regge model without unitarity corrections.
The explanation for this "paradox" is that the DL amplitude 
violations of s-unitarity are confined, even at super high energies,
to small b which does not contribute significantly to $\sigma_{tot}$. 
Note, though, that $\frac{\sigma_{el}}{\sigma_{tot}}$ grows in DL like 
$s^{\epsilon}$ whereas in GLM its growth is continuosly being moderated 
with increasing s (see table in {\bf 5.3}). 
The DL model predicts that $S^2$ 
is identical to unity or very close to it in the DL high-$t$ model where a 
weak $\pom\pom$ cut is added. The need for survival probabilities so as 
to reproduce the the experimental soft SD cross section values and the 
hard di-jets rates,
is the most compelling evidence 
in support of unitarization at presently available energies. 
As such, the study of high energy soft and hard diffraction serves as a 
unique probe substantiating the importance of s-channel unitarity in the 
analysis of high energy scattering.
\subsection{$S^2$ in unitarized models}
Most, but not all, of the unitarized models dealing with LHC 
$S^2$ predictions have roughly the same $S^2$ values. 
This calls for some clarifications. The first part 
of our discussion centers on the correlated investigation of two problems: 
\newline 
1) How uniform are the output predictions
of different unitarization procedures? 
\newline
2) How sensitive are the eikonal calculations to the details of the
eikonal model they use? 

We start with two non eikonal models which have 
contradictory predictions. 

The first is a model suggested by Troshin and Tyurin\cite{Troshin:2004gh}. In this 
model the single channel unitarity constraint (Eq.(\ref{2.13})) is 
enforced with an asymptotic bound where $G_{in}=0$ and $|a_{el}|=2$ 
i.e. asymptotically, $\sigma_{tot}=\sigma_{el}$ and $P^S(s,b)=1$.
The parameters of the model are set so as to obtain a "normal" 
survival probability monotonically decreasing with energy up to about 
$2500\,GeV$ where it changes its behavior and rises monotonically to its 
asymptotic limit of 1. Beside the fact that the model has a legitimate but 
non appealing asymptotics, its main deficiency is that it suggests a 
dramatic change in the systematics of $S^2$ without being able to offer 
any experimental signature to support this claim.
Regardless of this criticism, this is a good example of a proper unitarity 
model whose results are profoundly different from the eikonal model 
predictions. 

Another non eikonal procedure is Goulianos flux renormalization 
model\cite{Dino}. This is a phenomenological model which formally 
does not enforce unitarity, but rather, a bound of 
unity on the Pomeron flux in diffractive processes. 
Note that, the Pomeron flux is not uniquely defined so this should 
be regarded as an ad hoc parametrization. 
Nevertheless, it has scored an impressive success in reproducing the  
soft and hard diffractive data in the ISR-Tevatron range. 
The implied survival probabilities of this procedure are compatible with 
GLM and KKMR. 
However, the model predicts   
suppression factors for the diffractive channels which 
are $t$-independent and, thus, b-independent. 
The result is that, even though the output diffractive 
cross section is properly reduced relative to its input,
there is no change of the output profile from its 
input Gaussian form. Consequently, the Pumplin bound is violated. 
We are informed that Goulianos plans to improve his model by 
eikonalizing the output of his present model.

As noted, there are a few eikonal models on the 
market\cite{Block:2001ru,Eboli:1999dd,Eboli:2001fz,Frankfurt:2004kn,Frankfurt:2004ti,Petrov:2003yt,Bialas:2002rt,Bialas:2003xe}, 
and their predictions are compatible with GLM and KKMR.
Reconsidering the procedure of these calculations, their compatibility 
is not surprising once we translate their input to a GLM format. 
The GLM eikonal $S^2$ calculation has two input  
sectors in either a single or a two channel version. 
They are the soft $\nu$ and 
$R^2$, and the hard radius ${R^H}^2$. Since the soft input is based on 
a fit of the soft scattering data base, the potential variance in the 
soft parameters is relatively small. 
The input hard radius is 
obtained from either the HERA data or a theoretical 
calculation, be it a pQCD diagram or a Regge model. All in all, 
this is a reasonably stable input. 
In this context, it is interesting to discuss the eikonal model of 
Block and Halzen\cite{Block:2001ru}, where the calculated 
survival probabilities for Higgs production through W-W fusion are 
seemingly too high, 
$S^2(540)\,=\,27\%,\,S^2(1800)\,=\,21\%$ and $S^2(14000)\,=\,13\%$.
Even though, Higgs production is a CD process, 
the above $S^2$ values are in agreement with the KKMR calculations  
of $S_{DD}^2$ with a relatively high ${R^H}^2\,=\,11\,GeV^{-2}$. 
In a proper $S_{CD}^2$ calculation, these high $S^2$ values correspond 
to an even higher  
${R^H}^2\,\simeq\,20\,GeV^{-2}$, which is far too high as an estimate of 
the hard radius of $WW\rightarrow H$. A possible interpretation of 
Block-Halzen results is to associate them 
with a soft, rather than a hard, LRG CD process. 
This would couple with the non screened interpretation 
of CD Higgs through the soft CEM model\cite{Eboli:1999dd,Eboli:2001fz}, which 
predicts very high $S^2$ values. Since the CEM model is not screened we
may, as well, assign a survival probability to its output result. This 
translates into $S_{CD}^2\,=\,S_{BH}^2 S_{CEM}^2$, 
providing rather reasonable one channel predictions, 
$S_{CD}^2(540)=18.9\%$ and $S_{CD}^2(1800)=7.2\%$.

Obviously, each of the eikonal models, quoted above has its 
own particular presentation and emphasis. They do, however, have 
compatible results reflecting the observation that their input translates 
into similar values of $\nu,\,R^2$ and ${R^H}^2$.
\subsection{Compatibility between HERA and the Tevatron di-jets data}
Much attention has been given recently to the compatibility  
between the Tevatron and HERA DIS GJJ data. The starting point made by 
KKMR and CDF is that rather than depend on 
a p.d.f. input to calculate $F_{gap}$, we may use,
the GJJ di-jets diffractive structure function, $F_{jj}^D$,
inferred from HERA DIS data\cite{Derrick:1993xh,Derrick:1995wv,Derrick:1995tw,Derrick:1995pb,Breitweg:1998gc,Ahmed:1994nw,Ahmed:1995ns,Adloff:1997sc,schilling,Savin:2002eg} and associate it with the 
$F_{jj}^D$ derived from the Tevatron GJJ data. 
As it stands, this 
procedure ignores the role of the survival probability. 
Consequently, $F_{jj}^D$ obtained from the Tevatron  
is an order of magnitude smaller than the HERA  
output\cite{Abe:1994de,Abe:1997ie,Abe:1998ip,Affolder:2000vb,Affolder:2000hd,Dino,Khoze:2000cy,Khoze:2000wk,Khoze:2000jm,Kaidalov:2001iz,Kaidalov:2003ys}. 
This result led to speculations about a possible breaking of QCD 
or Regge factorization or both. 
Once the Tevatron di-jets diffractive structure function 
is rescaled by the appropriate survival probability, the 
compatibility between the 
Tevatron and HERA DIS diffractive data is attained. 
The conclusion of this 
analysis is that  the breaking of factorization is attributed to the soft   
re-scatterings of the the colliding projectiles. Additional hard 
contribution to the factorization breaking  due to gluon radiation is 
suppressed by the Sudakov factor included in the 
pQCD calculation (see {\bf 4.3}).
 
One should note, though, that the H1 determination\cite{Derrick:1993xh,Derrick:1995wv,Derrick:1995tw,Derrick:1995pb,Breitweg:1998gc,Ahmed:1994nw,Ahmed:1995ns,Adloff:1997sc,schilling,Savin:2002eg} of 
$F_{jj}^D$ 
is not unique. 
Arneodo\cite{Arneodo} has suggested a different $F_{jj}^D$ output based on 
HERA di-jets data which has a different normalization and $\beta$ 
dependences.
Should this be verified, there might well be a 
need to revise the KKMR calculations. 
 
The evolution of HERA $F_{jj}^D$ from high $Q^2$ DIS to $Q^2=0$ di-jets 
photoproduction has raised additional concern with regard to the validity 
of the factorization theorems\cite{Abramowicz:2004dt,gutsche}.
This is a complicated analysis since one has to be 
careful on two critical elements of the calculations:
\newline
1) The determination of the ratio between direct and resolved 
exchanged photon (real or virtual). This is a crucial element of the 
theoretical 
calculation since survival probability is applicable only to the 
resolved photon component. For very high $Q^2$ data  
the hard scattering process with the target partons is 
direct. At $Q^2=0$ there is a significant resolved photon 
contribution. 
\newline
2) For di-jets production there is a big difference between the LO 
and the NLO pQCD calculated cross sections\cite{Klasen:2004qr,Klasen:2004ct,Kaidalov:2003xf}. Since the HERA 
analysis compares the pQCD calculation with the di-jets measured cross 
section the normalization and shape of the theoretical input is most 
crucial in the experimental comparison between the high $Q^2$ and 
$Q^2\,=\,0$ data.
\newline 
On the basis of a NLO calculation,  
Klasen and Kramer\cite{Klasen:2004qr,Klasen:2004ct} conclude that they can 
reproduce the photoproduction data with $S^2\,=0.34$, 
applied to the resolved sector.
This survival probability is in agreement with KKMR and GLM 
calculations. 
\newline
Regardless of the above, preliminary photoproduction GJJ HERA 
data\cite{Abramowicz:2004dt,gutsche} suggest that both the direct 
and resolved photon sectors are suppressed at $Q^2=0$. 
A verification of this observation has severe consequences for our 
understanding of the evolution of the diffractive structure function from 
DIS to photoproduction. It does not directly relate, though, to the issue 
of soft survival probability which apply, per definition, only to the 
resolved photon sector. The suggested effect in the direct photon sector  
should, obviously be subject to a good 
measure of caution before being substantiated by further independent 
analysis. 

\subsection{Diffraction at energies above the LHC}
We end with a table showing the GLM two channel predictions for energies 
including the LHC, and up to the top Cosmic Rays energies. 
\begin{center}
\begin{tabular}{|r|r|r|r|r|r|r|}
\hline
\multicolumn{1}{|c}{\rule[-2ex]{0ex}{5ex}$\sqrt{s}\,[GeV]$} &
\multicolumn{1}{|c}{$\sigma_{tot}^{DL}\,[mb]$} &
\multicolumn{1}{|c}{$\sigma_{tot}^{GLM}\,[mb]$} &
\multicolumn{1}{|c}{$\sigma_{el}^{GLM}\,[mb]$} &
\multicolumn{1}{|c}{$\sigma_{sd}^{GLM}\,[mb]$} &
\multicolumn{1}{|c}{$B_{el}^{GLM}\,[GeV^{-2}]$} & 
\multicolumn{1}{|c|}{${S_{CD}^{GLM}}^2$} \\
\hline
540\,\,\,\,&60.1\,\,\,\,&62.0\,\,\,\,\,&12.3\,\,\,\,\,\,&8.7\,\,\,\,\,\,&
14.9\,\,\,\,\,\,\,\,\,\,\,\,\,&0.066\,\,\\
1800\,\,\,\,&72.9\,\,\,\,&74.9\,\,\,\,\,&15.9\,\,\,\,\,\,&10.0\,\,\,\,\,\,&
16.8\,\,\,\,\,\,\,\,\,\,\,\,\,&0.055\,\,\\
14000\,\,\,\,&101.5\,\,\,\,&103.8\,\,\,\,\,&24.5\,\,\,\,\,\,&12.0\,\,\,\,\,\,&
20.5\,\,\,\,\,\,\,\,\,\,\,\,\,&0.036\,\,\\
30000\,\,\,\,&114.8\,\,\,\,&116.3\,\,\,\,\,&28.6\,\,\,\,\,\,&12.7\,\,\,\,\,\,&
22.0\,\,\,\,\,\,\,\,\,\,\,\,\,&0.029\,\,\\
60000\,\,\,\,&128.4\,\,\,\,&128.7\,\,\,\,\,&32.8\,\,\,\,\,\,&13.2\,\,\,\,\,\,&
23.4\,\,\,\,\,\,\,\,\,\,\,\,\,&0.023\,\,\\
90000\,\,\,\,&137.2\,\,\,\,&136.5\,\,\,\,\,&35.6\,\,\,\,\,\,&13.5\,\,\,\,\,\,&
24.3\,\,\,\,\,\,\,\,\,\,\,\,\,&0.019\,\,\\
120000\,\,\,\,&143.6\,\,\,\,&142.2\,\,\,\,\,&37.6\,\,\,\,\,\,&13.7\,\,\,\,\,\,&
24.9\,\,\,\,\,\,\,\,\,\,\,\,\,&0.017\,\,\\
\hline
\end{tabular}
\end{center}
The, somewhat, surprizing observation is that the 
GLM calculated total cross sections are 
compatible with the DL simple Regge predictions 
all over the above energy range. This is a
reflection of the fact that even at exceedingly high energies
unitarization reduces the elastic amplitude at small enough b values to be
relatively insensitive to the calculation of $\sigma_{tot}$. On the
other hand, we see that $\sigma_{el}$ becomes more moderate in its energy
dependence and ${\sigma_{el}}/{\sigma_{tot}}$ which is $23.6\%$ at
the LHC is no more than $26.4\%$ at the highest Cosmic Rays
energy, $120\,TeV$. The implication of this observation is that the 
nucleon profile becomes darker at a very slow rate and is grey 
(well below the black disc bound) even 
at the highest energy at which we can hope for a measurment. A check of 
our results at the Planck scale 
shows $\sigma_{tot}\,=\,1010\,mb$ and the profile to be 
entirely black. i.e., 
$\frac{\sigma_{el}}{\sigma_{tot}}\,=\,\frac{1}{2}$.
$\sigma_{sd}$ is even more moderate in its very slow
rise with energy. 
The diminishing rates 
for soft and hard diffraction at exceedingly high energies are a
consequence of the monotonic reduction in the values of $S^2$  
with a Planck scale limit of $S^2\,=\,0$. This
picture is bound to have its effect on Cosmic Rays studies. 

Our LHC predictions are compatible with KMR.
Note, though, that: 
i) $\sigma_{sd}^{GLM}$ is rising slowly with $s$ gaining $20\%$ from the 
Tevatron to LHC. KMR has a much faster rise with energy, where, 
$\sigma_{sd}^{KMR}$ is gaining $77\%-92\%$ over the same energy interval. 
ii) At the LHC $B_{el}^{GLM}\,=\,20.5\,GeV^{-2}$, 
to be compared with a DL slope of $19\,GeV^{-2}$ and a KMR slope of 
$22\,GeV^{-2}$. 
The GLM $30\,TeV$ cross
sections are compatible with Block-Halzen.

\section {Acknowledgements}
We are very thankful to our colleagues Valery Khoze, Alan Martin, Misha 
Ryskin and Leif L\"onnblad,  
who generously contributed to {\bf Section 4} and the {\bf Appendix}.
Needless to say, they bear no responsibility for the rest of this review.

\section*{Appendix: Monte Carlo modeling of gap survival}
The following was contributed by Leif L\"onnblad and is presented without
any editing.

An alternative approach to gap survival and factorization breaking is
to implement multiple interactions in Monte Carlo event generators.
These models are typically based on the eikonalization of the partonic
cross section in hadronic collisions and can be combined with any hard
sub process to describe the additional production of hadrons due to
secondary partonic scatterings. Some of these programs, such as
\pythia\cite{Sjostrand:1987su,Sjostrand:2000wi} and
\herwig/\jimmy\cite{Corcella:2000bw,Butterworth:1996zw,Jimmy}, are
described in some detail elsewhere in these proceedings\cite{Buttar}.
Common for all these models is that they include exact kinematics and
flavour conservation, which introduces some non-trivial effects and
makes the multiple scatterings process-dependent. Also, the
predictions of the models are very sensitive to the cutoff used to
regularize the partonic cross section and to the assumptions made
about the distribution of partons in impact parameter space.
Nevertheless, the models are quite successful in describing sensitive
final-state observables such as multiplicity distributions and
jet-pedestal effects \cite{Buttar}. In particular this is true for the
model in \pythia which has been successfully tuned to Tevatron
data\footnote{Note that the model in \pythia has recently been revised
  \cite{}. However, the reproduction of Tevatron data is not as good
  for the revised model.} by Rick Field \cite{RickTuneA}, the
so-called \textit{CDF tune A}.

The \pythia model does not make any prediction for the energy
dependence of the total cross section - rather this is an input to the
model used to obtain the distribution in the number of multiple
interactions. \pythia can, however, make predictions for gap survival
probabilities. This was first done for Higgs production via W-fusion
\cite{Dokshitzer:1991he}, and amounts to simply counting the fraction
of events which do not have any additional scatterings besides the
W-fusion process. The basic assumption is that any additional partonic
scattering would involve a colour exchange which would destroy any
rapidity gap introduced by W-fusion process. Since \pythia produces
complete events, these can also be directly analyzed with the proper
experimental cuts. A similar estimate was obtained for the gaps
between jets process, both for the Tevatron and HERA
case\cite{Cox:1999dw}.

Recently, \pythia was used to estimate gap survival probabilities also
for the case of central exclusive Higgs production
\cite{Lonnblad:2004zp}. As in the case of gaps between jets, the
actual signal process is not implemented in \pythia, so direct
analysis with proper experimental cuts was not possible. Instead a
similar hard sub process was used (standard inclusive Higgs production
via gluon fusion in this case) and the fraction of events without
additional secondary partonic scatterings was identified with the gap
survival probability. Using the \textit{CDF tune A} the gap survival
probability was estimated to be 0.040 for the Tevatron and 0.026 for
the LHC. This is remarkably close both to the values used in
\cite{Khoze:2001xm} obtained in the KKMR model \cite{Kaidalov:2001iz},
and to the GLM values presented in section \ref{sec:S2pred} especially
the two-channel ones obtained in \cite{Gotsman}.

\newcommand{\refbrake}{\\\hspace*{2mm}}
\bibliographystyle{heralhc} 
{\raggedright\itemsep -1mm
\bibliography{LRG05}
}
\end{document}